%% file: paper_final.tex
\pdfoutput=1
\newif\ifarxiv
\arxivfalse
\def\paperversiondraft{draft}

\ifarxiv
  \documentclass[sigplan,screen,nonacm]{acmart}
\else
  \documentclass[sigplan,screen]{acmart}
\fi

\def\acmversionanonymous{anonymous}
\def\acmversionjournal{journal}

\makeatletter
\if@ACM@anonymous
  \def\acmversion{anonymous}
\else
  \def\acmversion{journal}
\fi
\makeatother

\usepackage{colortbl}

\usepackage[utf8]{inputenc}
\usepackage[T1]{fontenc}
\usepackage{microtype}
\usepackage{listings}
\usepackage{makecell}
\usepackage{xargs}
\usepackage{lipsum}
\usepackage{xparse}
\usepackage{xifthen, xstring}
\usepackage{xspace}
\usepackage{marginnote}
\usepackage{etoolbox}
\usepackage[acronym,shortcuts]{glossaries}
\usepackage{amsmath}
\usepackage{thmtools} %
\usepackage{algorithm}
\usepackage[noend]{algpseudocode}
\usepackage{hyphenat}
\usepackage[shortcuts]{extdash}

\input{tex/setup.tex}

\input{tex/acm.tex}

\makeatletter
\ifcsdef{MintedExecutable}
{
  \newminted[mlir]{tools/lexers/MLIRLexer.py:MLIRLexerOnlyOps}{mathescape}
  \newminted[xdsl]{tools/lexers/MLIRLexer.py:MLIRLexer}{mathescape, style=murphy}
  \newminted[lean4]{tools/lexers/Lean4Lexer.py:Lean4Lexer}{mathescape}
  \newmintinline[mlirinline]{tools/lexers/MLIRLexer.py:MLIRLexerOnlyOps}{mathescape}
  \newmintinline[xdslinline]{tools/lexers/MLIRLexer.py:MLIRLexer}{mathescape, style=murphy}
}
{
  \ifcsdef{minted@optlistcl@quote}
  {
    \newminted[mlir]{tools/lexers/MLIRLexer.py:MLIRLexerOnlyOps}{customlexer, mathescape}
    \newminted[xdsl]{tools/lexers/MLIRLexer.py:MLIRLexer}{customlexer, mathescape, style=murphy}
    \newminted[lean4]{tools/lexers/Lean4Lexer.py:Lean4Lexer}{customlexer, mathescape}
    \newmintinline[mlirinline]{tools/lexers/MLIRLexer.py:MLIRLexerOnlyOps}{customlexer, mathescape}
    \newmintinline[xdslinline]{tools/lexers/MLIRLexer.py:MLIRLexer}{customlexer, mathescape, style=murphy}
  }
  {
    \newminted[mlir]{tools/lexers/MLIRLexer.py:MLIRLexerOnlyOps -x}{mathescape}
    \newminted[xdsl]{tools/lexers/MLIRLexer.py:MLIRLexer -x}{mathescape, style=murphy}
    \newminted[lean4]{tools/lexers/Lean4Lexer.py:Lean4Lexer -x}{mathescape}
    \newmintinline[mlirinline]{tools/lexers/MLIRLexer.py:MLIRLexerOnlyOps -x}{mathescape}
    \newmintinline[xdslinline]{tools/lexers/MLIRLexer.py:MLIRLexer -x}{mathescape, style=murphy}
  }
}
\makeatother

\newacronym{ir}{IR}{Intermediate Representation}
\newacronym{wse}{WSE}{Wafer-Scale Engine}
\newacronym{csl}{CSL}{Cerebras Software Language}
\newacronym{dsd}{DSD}{Data Structure Descriptor}
\newacronym{pe}{PE}{Processing Element}
\newacronym{hpc}{HPC}{High-Performance Computing}
\newacronym{dsl}{DSL}{Domain-Specific Language}

\graphicspath{{./images/}}

\makeatletter
\newcommand\requiredelimiter[2][########]{%
  \ifdefined#2%
    \def\@temp{\def#2#1}%
    \expandafter\@temp\expandafter{#2}%
  \else
    \@latex@error{\noexpand#2undefined}\@ehc
  \fi
}
\@onlypreamble\requiredelimiter
\makeatother

\newcommand\newdelimitedcommand[2]{
\expandafter\newcommand\csname #1\endcsname{#2}
\expandafter\requiredelimiter
\csname #1 \endcsname
}

\newdelimitedcommand{toolname}{Tool}

\usepackage{booktabs}

\usepackage[verbose]{newunicodechar}
\newunicodechar{₁}{\ensuremath{_1}}
\newunicodechar{₂}{\ensuremath{_2}}
\newunicodechar{∀}{\ensuremath{\forall}}
\newunicodechar{α}{\ensuremath{\alpha}}
\newunicodechar{β}{\ensuremath{\beta}}

\DeclareRobustCommand{\circled}[2][]{%
    \ifthenelse{\isempty{#1}}%
        {\circledbase{pairedOneLightBlue}{#2}}%
        {\autoref{#1}: \hyperref[#1]{\circledbase{pairedOneLightBlue}{#2}}}%
}

\begin{document}

\copyrightyear{2026}
\acmYear{2026}
\setcopyright{cc}
\setcctype{by}
\acmConference[ASPLOS '26]{Proceedings of the 31st ACM International Conference on Architectural Support for Programming Languages and Operating Systems, Volume 2}{March 22--26, 2026}{Pittsburgh, PA, USA}
\acmBooktitle{Proceedings of the 31st ACM International Conference on Architectural Support for Programming Languages and Operating Systems, Volume 2 (ASPLOS '26), March 22--26, 2026, Pittsburgh, PA, USA}
\acmPrice{}
\acmDOI{10.1145/3779212.3790124}
\acmISBN{979-8-4007-2359-9/2026/03}

\title{An MLIR Lowering Pipeline for Stencils at Wafer-Scale}

\author{Nicolai Stawinoga}
\authornote{Both authors contributed equally to this work}
\authornote{Affiliation at the time of contribution}
\orcid{0000-0002-3806-2691}
\affiliation{
  \institution{Technische Universität Berlin}  
  \city{Berlin}  
  \postcode{10587}
  \country{Germany}
}
\email{nicolai@aes.tu-berlin.de}         

\author{David Katz}
\authornotemark[1]
\orcid{0009-0003-7387-6169}
\affiliation{
  \department{EPCC}
  \institution{The University of Edinburgh}  
  \city{Edinburgh}  
  \country{United Kingdom}
}
\email{d.kacs@sms.ed.ac.uk}         

\author{Anton Lydike}
\orcid{0009-0001-9389-8512}             
\affiliation{
  \department{School of Informatics}
  \institution{The University of Edinburgh}
  \city{Edinburgh}    
  \country{United Kingdom}
}
\email{anton.lydike@ed.ac.uk}

\author{Justs Zarins}
\orcid{0000-0002-0633-1404}             
\affiliation{
  \department{EPCC}             
  \institution{The University of Edinburgh}             
  \city{Edinburgh}  
  \country{United Kingdom}
}
\email{j.zarins@epcc.ed.ac.uk}

\author{Nick Brown}
\orcid{0000-0003-2925-7275}             
\affiliation{
  \department{EPCC}             
  \institution{The University of Edinburgh}           
  \city{Edinburgh}  
  \country{United Kingdom}
}
\email{nick.brown@ed.ac.uk}         

\author{George Bisbas}
\authornotemark[2]
\orcid{0000-0002-1519-1028}             
\affiliation{  
  \institution{Imperial College London}         
  \city{London}  
  \country{United Kingdom}
}
\email{georgios.a.bisbas@gmail.com}        

\author{Tobias Grosser}
\orcid{0000-0003-3874-6003}             
\affiliation{
  \institution{University of Cambridge}           
  \city{Cambridge}  
  \country{United Kingdom}
}
\email{tobias.grosser@cst.cam.ac.uk}         

\renewcommand{\shortauthors}{Stawinoga, Katz et al.}

\begin{abstract}
The Cerebras Wafer-Scale Engine (WSE) delivers performance at an unprecedented scale of over 900,000 compute units, all connected via a single-wafer on-chip interconnect. Initially designed for AI, the WSE architecture is also well-suited for High Performance Computing (HPC). However, its distributed asynchronous programming model diverges significantly from the simple sequential or bulk-synchronous programs that one would typically derive for a given mathematical program description. Targeting the WSE requires a bespoke re-implementation when porting existing code. The absence of WSE support in compilers such as MLIR, meant that there was little hope for automating this process.

Stencils are ubiquitous in HPC, and in this paper we explore the hypothesis that domain specific information about stencils can be leveraged by the compiler to automatically target the WSE without requiring application-level code changes. We present a compiler pipeline that transforms stencil-based kernels into highly optimized CSL code for the WSE, bridging the semantic gap between the mathematical representation of the problem and the WSE’s asynchronous execution model. Based upon five benchmarks across three HPC programming technologies, running on both the Cerebras WSE2 and WSE3, our approach delivers comparable, if not slightly better, performance than manually optimized code. Furthermore, without requiring any application level code changes, performance on the WSE3 is around 14$\times$ faster than 128 Nvidia A100 GPUs and 20$\times$ faster than 128 nodes of a CPU-based Cray-EX supercomputer when using our approach.

\end{abstract}

\ifx\acmversion\acmversionanonymous
\settopmatter{printacmref=false} %
\renewcommand\footnotetextcopyrightpermission[1]{} %
\fi
\ifx\acmversion\acmversionjournal


\fi

\maketitle

\nocite{brown2021accelerating}
\nocite{brown2022exploring}
\nocite{de2020transformations}
\nocite{gianinazzi2025spada}
\nocite{jacquelin2022scalable}
\nocite{ltaief2023scaling}
\nocite{orenes2023wafer}
\nocite{perez2025breaking}
\nocite{santos2024breaking}
\nocite{tramm2024efficient}
\nocite{trifan2022intelligent}
\nocite{woo2022disruptive}
\nocite{zvyagin2023genslms}

\section{Introduction}

The Cerebras \ac{wse}~\cite{lie2023cerebras} is a \emph{systolic array}~\cite{kung1979systolic} spanning a full wafer of 900,000 independent interconnected \acp{pe}
on the \ac{wse}3. Originally designed as a novel class of AI accelerator, and now in its third architectural generation, the Cerebras \ac{wse} has been applied to a variety of impactful HPC applications~\cite{trifan2022intelligent,zvyagin2023genslms,woo2022disruptive,jacquelin2022scalable,ltaief2023scaling,santos2024breaking,tramm2024efficient,orenes2023wafer,perez2025breaking}. Indeed \citet{ltaief2023scaling} demonstrated that due to the large amount of distributed SRAM, the WSE is capable of delivering performance where other architectures hit memory-bound bottlenecks. Driven by the potential for very high performance, there is significant interest in using the WSE for HPC workloads, however very few people are actually using the architecture and this is due to the major blocker being the effort required to port codes.

Although a significant time is required to rewrite codes in Cerebras's bespoke general-purpose CSL programming language, the programmer's first challenge is  the algorithmic transformation required to suit the WSE. This recasting of an algorithm requires substantial time and expertize on behalf of the programmer, often involving architecture-specific techniques. Consequently, whilst there have been several heroic efforts to bring simple kernels or benchmarks to the WSE, the technology is yet to enjoy widespread use or adoption in HPC due to these programming challenges.

In this paper we present a compiler flow, based on MLIR, that enables codes to run on the WSE without modification. Our overarching research hypothesis is that it is possible to solve the fundamental programmability and performance challenges on esoteric architectures, such as the WSE, by using compiler automation driven by domain specific abstractions present in user code. Targeting three front-ends across two scientific domains our proposition is that software tooling, such as MLIR, that was initially designed with AI workloads in mind can be leveraged to solve some of the fundamental challenges around exploiting AI-based hardware for High Performance Computing (HPC).

\vspace{.5em}
\noindent
The contributions of this paper are:
\begin{itemize}
  \item A discussion of the WSE's highly asynchronous execution model and how it affects program representation when targeted by code generation (Section~\ref{sec:async}).
  \item A set of intermediate domain-specific and domain-agnostic dialects to convert stencil expressions to CSL code via our low-level \texttt{\small csl-ir} dialect that re-implements a large subset of CSL (Section~\ref{sec:dialects}).
  \item A strategy of lowering transformations to generate the \texttt{\small csl-ir} dialect from higher level MLIR abstractions (Section~\ref{sec:cs-compiler}).
  \item An evaluation of our stencil lowering pipeline for the WSE2 and WSE3, ultimately demonstrating that our approach delivers slightly better performance than hand-written code supporting  multiple generations of the WSE -- likely the first performance comparison of the WSE2 and WSE3 at the time of writing (Section~\ref{sec:evaluation}).
\end{itemize}

\section{Motivation: The Cerebras WSE and its asynchronous execution model}
\label{sec:async}

\begin{figure}[htb]
    \centering
    {\includegraphics[width=\columnwidth]{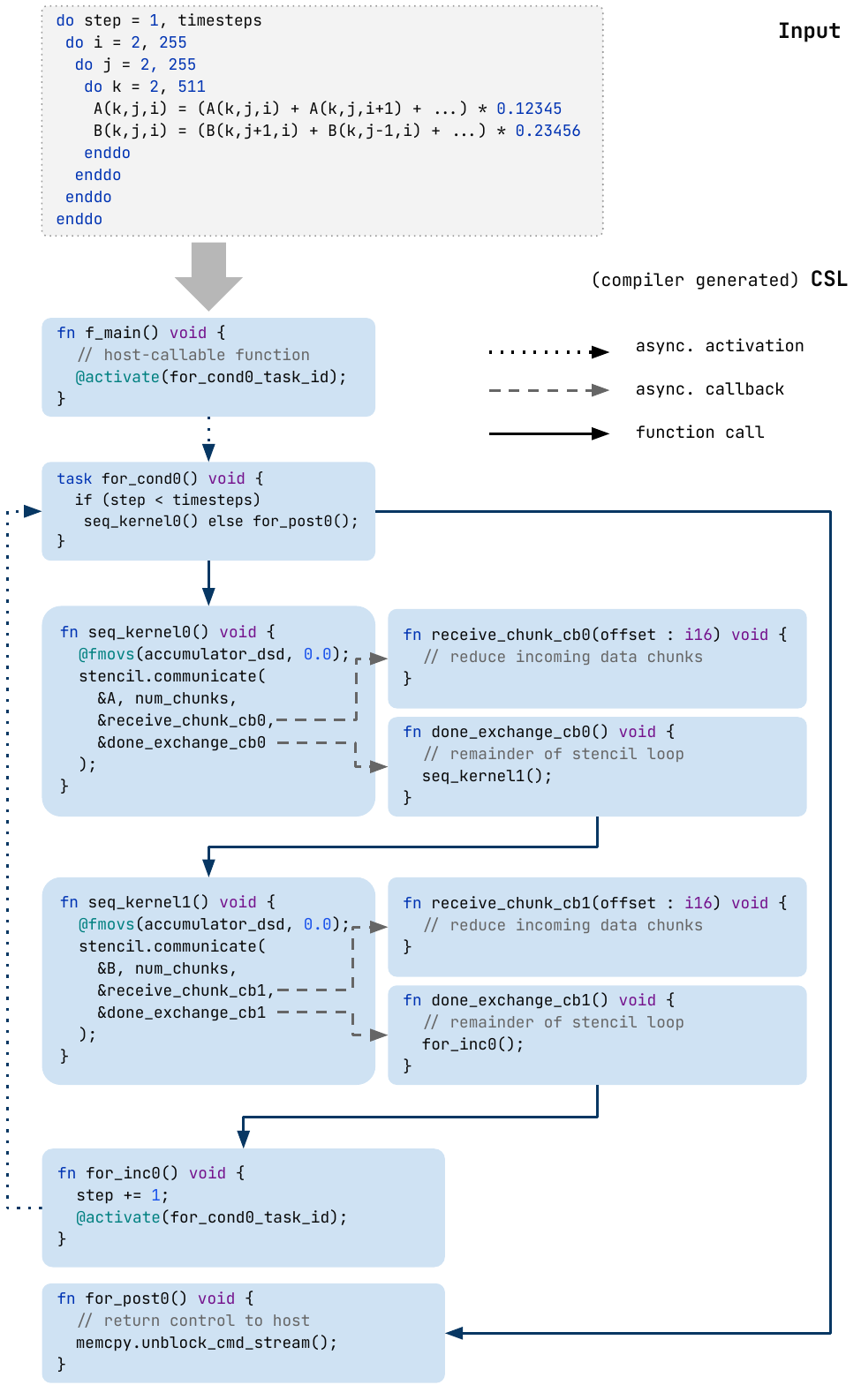}}
    \caption{A time-step loop surrounding asynchronous data exchanges in CSL.}
    \label{fig:time_step_loop}
\end{figure}

The Cerebras Wafer-Scale Engine (WSE) features a unique and highly-asynchronous execution model characterized by its compute architecture. It features a systolic array of up to 900,000 Processing Elements (PEs), where each comprises a computational core, router, program counter, clock, and 48KB of SRAM memory. PEs can be programmed individually and act fully independently of one another, managing their own memory to which they are able to perform a 128-bit memory read and a 64-bit memory write in each cycle. The impressive total 40GB (44GB) of on-chip memory for the WSE2 (WSE3) architecture leveraged in this work does not form a unified memory system, and no caches or memory hierarchy exists.

Data between PEs is shared explicitly via four bi-directional communication channels, connecting the routers of PEs to that of their immediate cardinal north, south, east, and west neighbors. An additional channel, called \emph{the ramp}, connects the router to the attached core. A 32-bit message, known as a \emph{wavelet}, can be sent between neighboring PEs in a single clock cycle in each direction, resulting in an aggregate peak fabric bandwidth of 214 Pb/s on the WSE3. Twenty-four virtual channels, known as \emph{colors}, provide an abstraction to the programmer for configuring routes and switches. When an asynchronous data transfer has completed, the PE can flag an internally bound \emph{task} as unblocked or as ready for activation.

In 2022 Cerebras released their Cerebras Software Language (CSL) SDK which enables direct programming of the WSE's architecture. CSL is syntactically based on Zig, featuring numerous extensions to explicitly support the WSE, such as tasks, wavelets, colors and Data Structure Descriptors (DSDs) which enable performing operations across arrays for efficiency. However, the exposing of the WSE's execution model in practice makes CSL programming appear fundamentally different from programming in Zig. Besides in-depth knowledge of the architecture, programmers must also fundamentally recast their algorithms moving from a form suitable to CPUs or GPUs to one fitting the WSE.

Figure~\ref{fig:time_step_loop} illustrates the representation of a time-step loop surrounding a naive Jacobian stencil program in Fortran (top) and CSL (bottom). The assumption is that the two stencil operations contained in the time-step require asynchronous communication between PEs. Therefore, although CSL provides various loop constructs, it is not possible to represent this program in a top-level loop, as CSL lacks a mechanism to achieve re-synchronization within a code block (e.g., \emph{await}/\emph{barrier} in single-threaded and multi-threaded contexts, respectively). Consequently, any control flow that surrounds asynchronous operations such as communication must be translated into tasks and callbacks. The programmer must therefore decompose the program into separate tasks which are then scheduled for execution based on data arrival. In languages designed for CPUs and conventional accelerators, an outer loop such as \emph{do step = 1, timesteps} is a simple language construct, however on the WSE this same algorithm must be recast as tasks driven by callbacks, due the communication required between iterations of the outer loop.

The call to \texttt{\small stencil.communicate()} in \emph{seq\_kernel0()} and \emph{seq\_kernel1()} of Figure~\ref{fig:time_step_loop} encapsulates boiler-plate code for
chunk-wise data exchanges and internally uses additional tasks. Practicalities of the architecture, specifically the limited amount of memory per PE, severely limit
the size of buffers and-so data must be exchanged in chunks. Consequently, this function accepts two user-provided callbacks (triggered indirectly as indicated by dashed arrows); the first that is triggered to process
each incoming chunk of data, and the second is executed once the entire exchange of data has completed. The algorithm's overall execution flow of is then continued
from the second callback, and this demonstrates the complexity faced by the programmer when they wish to port even a seemingly simple kernel to the WSE because
CSL directly exposes low-level architectural concepts, making the programming complex and susceptible to errors.

\subsection{An actor-based asynchronous execution model}
\label{sec:actor_exec_model}
The semantic gap in program representations depicted in Figure~\ref{fig:time_step_loop} directly follows differences in the execution model of the underlying architecture. Unlike conventional data-parallel accelerators, the WSE's is primarily a task-parallel architecture in a manner resembling the \emph{actor} paradigm. Originally published under the title \emph{A universal modular ACTOR formalism for artificial intelligence} in 1973 \cite{hewitt1973universal} (over 50 years ago), the formalism predates hardware and software implementations, yet well describes the execution model of the WSE both on a hardware and software level.

On the hardware level, we conceive of \emph{PEs as hardware actors} in a static arrangement. A key characteristic of the actor model is that data is passed as messages between actors and on the WSE each PE contains a set of queues, acting as mailboxes, into which data is received. Actors also hold their own state, and on the WSE each PE acts independently and contains its own clock and private memory. There are two types of messages, or \emph{wavelets} that can be passed between hardware actors, \emph{data wavelets} that contain a data payload, and \emph{control wavelets} to instruct another actor to update its routing configuration. Depending on the PEs state messages are forwarded or consumed, which in some cases involves activating a software actor. PEs as hardware actors have a physical location, a concept not shared by software implementations of the actor model, and whilst there is library support to pass messages between arbitrary PEs, exploiting spatial proximity for neighbor to neighbor communication is crucial for performance in practice.

On a software level, we conceive of \emph{tasks as software actors}.
The code running inside each PE provides a software implementation of the actor model via tasks. There are three types of tasks provided by CSL, each of which with hardware support:
\begin{itemize}
\item \texttt{\small data\_task}s are message listeners triggered when receiving a message, or \emph{data wavelet}, in the form of data on the queue they are configured to listen to.
\item \texttt{\small control\_task}s are similar to data tasks but are triggered by \emph{control wavelets}, messages that do not carry a data payload.
\item \texttt{\small local\_task}s can only be triggered internally and are particularly useful as callbacks when completing asynchronous sending or receiving of data.
\end{itemize}

As program state is maintained by the PE, tasks have a shared program state, although separate state could theoretically be enforced by the programmer. Tasks as software actors do not exhibit true parallelism, as each PE is strictly single-threaded, but instead tasks provide an asynchronous execution model. Passing execution flow between tasks adds significant complexity, and is frequently referred to as \emph{`callback hell'} in literature \cite{edwards2009coherent}. The major challenge is that CSL exposes this problem to the programmer without providing higher-level language abstractions, such as \emph{futures} \cite{caromel2009asynchronous}, that could help address it. Re-structuring a program from its synchronous representation to the asynchronous model creates the \emph{continuation complexity problem} \cite{zamora2015continuation} of passing execution flow, and this was a challenge we had to address in our IR transformation as described in Section \ref{sec:cs-compiler}.

To summarize, much of the complexity involved in programming the WSE is due to having to map synchronous algorithms to the asynchronous nature of the WSE. It is important to note, however, that CSL is not formally an actor-based programming language, exposing the user to problems known from asynchronous actor-like programming, including the complexity of managing interactions in barrier-free parallelism, preventing race conditions, and more.

\subsection{Departing from classic dataflow architectures}

Whilst the WSE is a dataflow architecture, the coarse grained nature of the design means that it is very different from other, fine grained, dataflow architectures such as FPGAs and this provides unique challenges for distributed stencil computations and compilation. Unlike single units of logic that stream data in fine-grained dataflow architectures, each PE in the WSE contains a vectorized CPU that runs tasks and is capable of a much greater level of compute. Consequently, the pipelined approach to parallelism that is natural on fine grained architectures such as FPGAs is not applicable on the WSE, as workloads comprise both local and remote compute. Indeed, on the WSE the programmer must adopt a combination of parallelism and optimization techniques across both dataflow (e.g. communication between PEs involves streaming many individual pieces of data) and distributed memory (e.g. for stencils decomposing geometrically across PEs) architectures and this hybrid approach is very difficult.

As an example, leveraging shift buffers \cite{de2020transformations} as bespoke memory on FPGAs to implement highly efficient stencil computations is a well known technique \cite{brown2021accelerating}. However the nature of the WSE means that this approach is neither possible, as memory is local and private to PEs, nor optimal, as one must carefully design inter-PE communications. Instead, for efficient execution on the WSE one must adopt a different approach to decomposing their problem across this dataflow architecture and, as described in this section leveraging asynchronous tasks that interleave local computation and handling the arrival of remote data.

\section{Compiler ecosystems and stencils}

\subsubsection*{MLIR}
MLIR \cite{mlir} provides a set of reusable IR dialects and transformations between them; separate dialects can be mixed into the IR and manipulated independently. Similar to the more low-level LLVM-IR \cite{lattner2004llvm}, MLIR follows the Static Single Assignment (SSA) form. Reusing existing dialects and transformations provides much greater sharing of compiler infrastructure, significantly reducing the overall software effort in developing new compilers. Furthermore, MLIR is also a framework where compiler developers can create their own dialects and transformations. The pluggable nature of the framework allows the creation of deep compile chains with a wide range of reusable dialects, backends, or frontends.

\subsubsection*{xDSL}

MLIR imposes a steep learning curve on developers where they must, for example, understand LLVM concepts, work with the Tablegen format to describe dialects, and keep track of the fast-evolving MLIR repository. xDSL was developed as a Python-based compiler design toolkit that is 1-1 compatible with MLIR. Dialects are expressed in the IRDL~\cite{fehr2022irdl} format within Python classes, enabling rapid exploration and prototyping of MLIR concepts, with a more mature version then committed into the MLIR codebase once proven. The MPI dialect is one example of this approach, where it was first developed in xDSL \cite{bisbas2024shared} before being standardized into MLIR. As xDSL is 1-1 compatible with MLIR, the IR can freely pass between the two frameworks during compilation. We used xDSL to develop the transformations and optimizations described in this paper, and due to xDSL being based upon MLIR, we use the two terms interchangeably unless otherwise stated.

\subsubsection*{The stencil dialect}

\begin{listing}[h]
\begin{lstlisting}[language=Fortran, frame=none, numbers=left, basicstyle=\scriptsize]
do i = 2, 255
  do j = 2, 255
    do k = 2, 511
      data(k,j,i) = (data(k,j,i) + data(k,j,i+1)) * 0.12345
    enddo
  enddo
enddo
\end{lstlisting}
\caption{Sketch of Fortran stencil code example which adds values from two grid cells and multiplies result by a constant
\label{lst:fortran_stencil_example}}
\end{listing}

The \texttt{\small stencil} dialect was initially developed as part of the Open Earth Compiler \cite{gysi2021domain} and is now part of xDSL. A stencil is a geometric arrangement of a group of neighboring grid cells that, by using a numerical approximation routine, relate to a specific grid cell of interest. Listing \ref{lst:fortran_stencil_example} provides an example of a Fortran stencil code which multiplies the result of adding values from two grid cells by a constant, and Listing \ref{lst:mlir_stencil_b} (see Section~\ref{sec:cs-compiler}) sketches the corresponding SSA-based MLIR using the \texttt{\small stencil} and \texttt{\small arith} dialects to express this calculation. The nested loops at lines 1, 2 and 3 of Listing \ref{lst:fortran_stencil_example} have been transformed into the \mlirinline{stencil.apply} operation in the IR, where the \mlirinline{stencil.access} operations in Listing \ref{lst:mlir_stencil_b} correspond to accesses on the \emph{data} array at line 4 of the Fortran code. Within the \mlirinline{stencil.apply} operation the \texttt{\small stencil} dialect determines data accesses, and the \texttt{\small arith} dialect determines the computation. It should be noted that the \mlirinline{stencil.apply} operator is running across the entire grid, whose lower and upper bounds are determined by the types of the input and output fields, executing the body of the \mlirinline{stencil.apply} operation for every grid cell.

\subsubsection*{Stencil generation from PSyclone, Devito \& Flang}

PSyclone~\cite{adams2019lfric} is a climate and weather DSL enabling scientists to write their kernels in Fortran, while abstracting away the low-level details of parallelism. It is used in the Met Office's next-generation weather and ocean models \cite{melvin2017lfric,porter2018portable}. The PSyclone compiler analyzes the scientist's code and undertakes optimizations and transformations upon an internal IR, such as applying OpenMP or OpenACC at the loop level for threaded parallelism and GPU acceleration respectively.

Devito~\cite{devito-api,devitoTOMS2020,bisbas2025automated}, is an open-source Python-based DSL and compiler framework widely used in academia and industry. Based around a symbolic DSL in Sympy~\cite{sympy2017}, its purpose is to ease the development of HPC Finite-Difference (FD) PDE solvers. Devito is used in a range of industries, especially those focused on seismic and medical workloads.

It was highlighted in \citeauthor{bisbas2024shared}~\citep{bisbas2024shared} that whilst these two DSLs appear rather different, they share a lot of common ground for compiler optimizations. Consequently, \citeauthor{bisbas2024shared}~\cite{bisbas2024shared} integrated PSyclone and Devito with the \texttt{\small stencil} dialect via xDSL, the objective being to share compiler infrastructure. It was observed that by being in the MLIR ecosystem both DSLs benefited from optimization passes developed by the wider community, including the vendors, and indeed this integration resulted in performance benefits on CPUs and GPUs. These optimizations are further discussed in Section \ref{sec:optimization}. This motivated integration of the \texttt{\small stencil} dialect with the Flang Fortran compiler \cite{brown2023fortran}, which itself is built atop MLIR. Performance delivered by Flang fell significantly short of other compilers \cite{brown2024fully}, and by the addition of a compiler pass identifying and extracting stencils within Fortran code, it was demonstrated that one could close this gap \cite{brown2023fortran}.

It is these three frontends, PSyclone, Devito, and Flang, that we leverage in this work and use as benchmarks in Section \ref{sec:evaluation}. Indeed, \citeauthor{rodriguez2023stencil} developed a lowering from the \texttt{\small stencil} dialect to an efficient FPGA representation, demonstrating that no changes were needed in the front end languages or DSLs. We adopt same general design approach here in Sections \ref{sec:dialects} and \ref{sec:cs-compiler}, in this case transforming the \texttt{\small stencil} dialect to a form that suits the Cerebras WSE.

\section{Intermediate MLIR dialects for the WSE}

\label{sec:dialects}
\begin{figure}[htb]
    \centering
    {\includegraphics[width=\columnwidth]{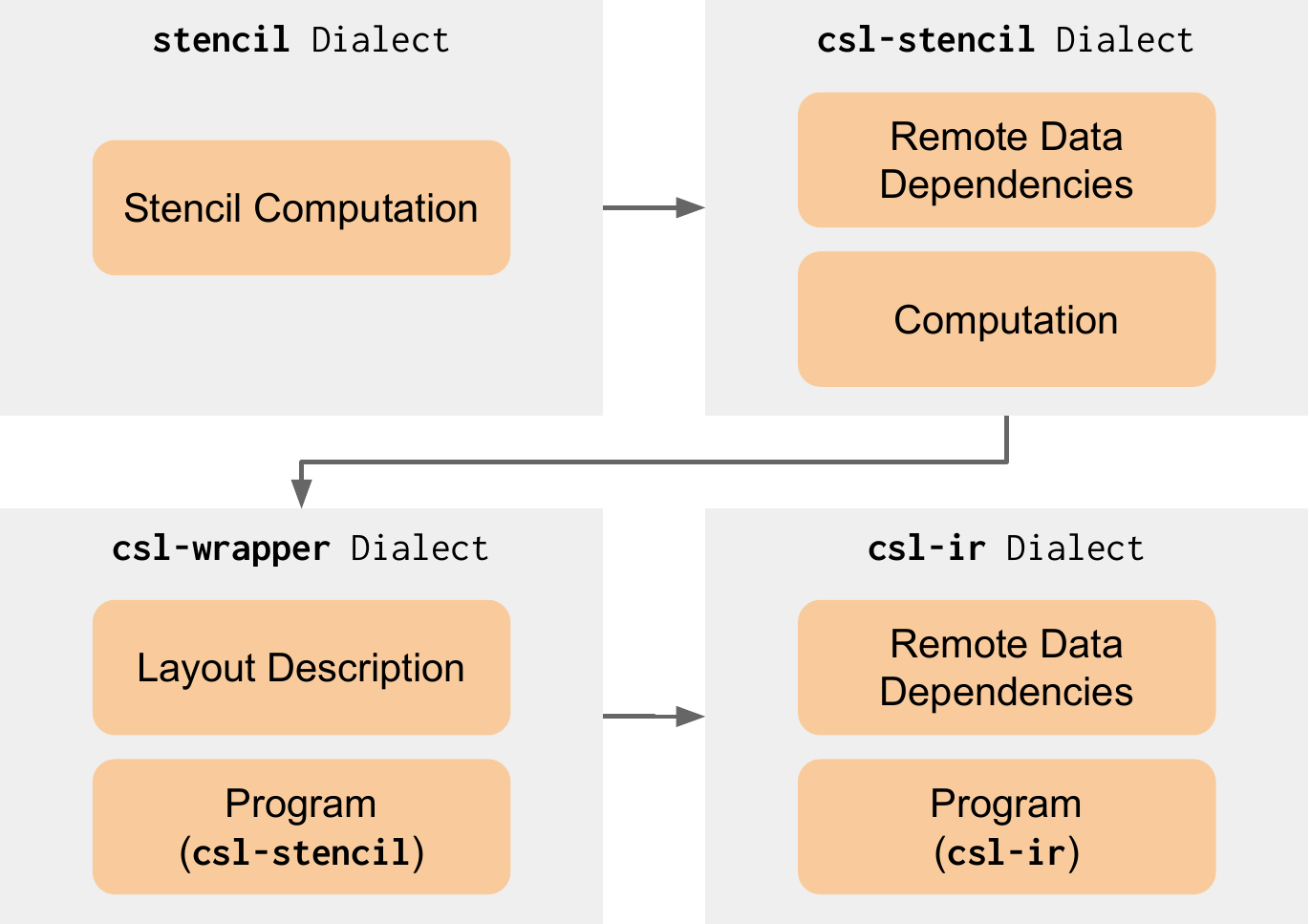}}
    \caption{Through the use of three WSE MLIR dialects, we incrementally move from the stencil mathematical representation to the actor-based WSE execution model.}
    \label{fig:new_dialects}
\end{figure}

There are two major groups of components comprising our MLIR based compiler flow for the WSE, the dialects and transformations that operate upon them. In this section we propose three new MLIR dialects for the Cerebras WSE, \texttt{\small csl-stencil}, \texttt{\small csl-wrapper}, and \texttt{\small csl-ir}. Figure~\ref{fig:new_dialects} provides an overview of how these integrate, where the existing \texttt{\small stencil} dialect is lowered to the \texttt{\small csl-stencil} dialect, which is then wrapped in the \texttt{\small csl-wrapper} dialect. Lastly, this is all transformed into operations in the \texttt{\small csl-ir} dialect.

\subsection{The csl-stencil dialect}

Stencils are lowered from the architecture-agnostic \texttt{\small stencil} dialect into the WSE-specific \texttt{\small csl-stencil} dialect, which explicitly specifies the communications and stencil computation operations. The \texttt{\small csl-stencil} dialect contains the following operations:

\begin{enumerate}
    \item \mlirinline{csl-stencil.prefetch} which describes the fetching of a single piece of remote data from its source into a local buffer.
    \item \mlirinline{csl-stencil.apply} which contains two regions, the first describes the processing of \emph{chunks} of data received from remote dependencies, and the second region describes compute operations on data held locally, which is done with respect to any pre-processing already applied by the first region.
    \item \mlirinline{csl-stencil.access} based upon an offset from the current index will access a value from the stencil, either directly if this is held locally or from an internal buffer if held remotely. As such, this op may eventually be lowered to either a local memory access or to an access to a buffer into which incoming remote data is received.
\end{enumerate}

Stencil computations by definition describe a reduction operation over a set of neighboring values (c.f. Listing~\ref{lst:fortran_stencil_example}). The \texttt{\small csl-stencil} dialect enables us to perform a two-fold partial reduction where applicable. First, by supporting the reduction of remote data separately from data held locally, via the \mlirinline{csl-stencil.apply} operation's first region. Secondly, by splitting communication into $n$ chunks of data and reducing remote data on a chunk-by-chunk basis. In this case the first region is executed $n$ times, but is able to immediately reduce incoming chunks of data into a single accumulator buffer, significantly lowering the overall memory requirements, of which each PE only provides a limited 48 kB. Supporting an optimization proposed by \cite{jacquelin2022scalable}, which was shown to be highly efficient on the WSE, the \mlirinline{csl-stencil.apply} operation also enables constants to be applied to incoming data, rather than these being located in the compute region. Subsequently in the pipeline these are mapped to CSL primitives that apply constants at zero overhead to incoming data, effectively interleaving communication and compute.

\subsection{The csl-wrapper dialect}

CSL programs involve two main components, the program itself as well as a layout file which is a metaprogram that determines placement and routing of programs across the WSE, along with compile-time constants. CSL adopts a staged compilation approach where the compiler executes this layout file and will then use this to specialize and compile programs for individual cores. At this stage, any code labeled \texttt{\small comptime} in the \ac{pe} programs is evaluated, parameters are lowered to constants, and the resulting PE program is aggressively optimized by the CSL SDK. If parameter values of \ac{pe} programs diverge, all encountered variants are compiled separately.
We provide the \texttt{\small csl-wrapper} dialect to handle this staged compilation approach, which is a domain-agnostic dialect that packages program-wide parameters, the layout metaprogram, and the PE program(s) together, alongside utility functionality to instantiate the program and metaprogram modules. It contains the following main operations:

\begin{enumerate}
    \item \mlirinline{csl-wrapper.module} defines a module with optional program parameters and two regions. The first region controls \emph{layout} across the WSE, and the second is for the program(s) to be mapped to PEs.
    \item \mlirinline{csl-wrapper.import} will import CSL-specific libraries and helper functions, such as the CSL \emph{memcpy} library which provides support for data transfers to and from the host.
\end{enumerate}

Layout files, which are written in CSL, tend to comprise a loop nest iterating over the \texttt{\small x} and \texttt{\small y} dimensions of the WSE and individually placing kernels onto the PEs. Given that we run the same program on each PE, the layout in our approach corresponds to a view of this inner loop nest.

\subsection{The \texttt{\small csl-ir} dialect}

The \texttt{\small csl-ir} dialect is a direct re-implementation of a large subset of the CSL programming language, where constructs present in the language are represented in the same manner in \texttt{\small csl-ir}. This is because, ultimately, we invoke a printer that generates CSL code from \texttt{\small csl-ir} intermediate representation. As such, all operations described above are eventually lowered into the \texttt{\small csl-ir} dialect and then from this CSL code is generated which is input to the Cerebras SDK compiler.

\section{Compiler transformations for the WSE}

\label{sec:cs-compiler}

\begin{figure*}[htb]
    \centering
    {\includegraphics[width=\textwidth]{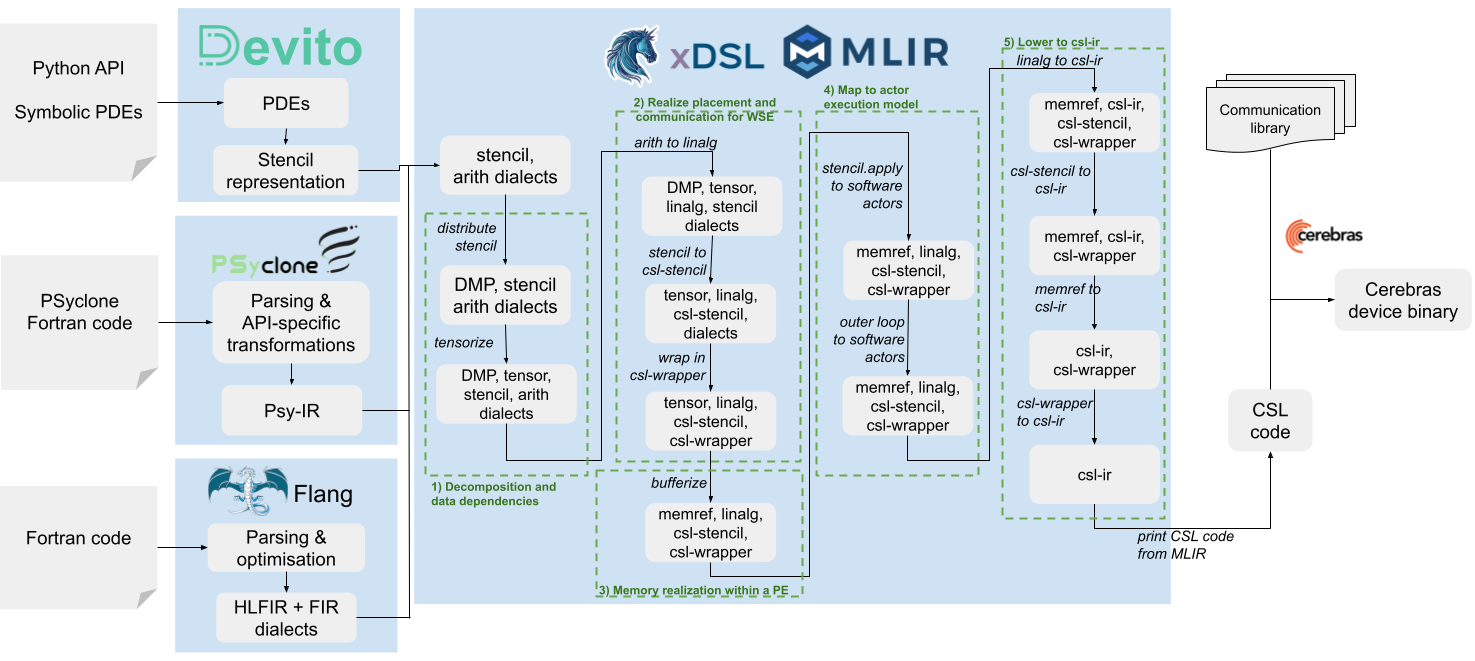}}
    \caption{Our staged compilation flow extends existing infrastructure to target novel hardware through multiple stages, allowing efficient code generation for the Cerebras WSE.}
    \label{fig:overall_flow}
\end{figure*}

The previous section presented both domain-specific and domain-agnostic intermediate dialects used to represent the structure of the IR at different stages during the transformation. Figure \ref{fig:overall_flow} sketches a detailed overall flow of the compilation pipeline with the main transformations that operate upon the IR where, starting from the Python or Fortran code provided as input to the Devito, PSyclone or Flang compilers, existing work \cite{bisbas2024shared, brown2023fortran} is leveraged to transform these into the \texttt{\small stencil} dialect within the xDSL framework.

This flow involves executing a pipeline of transformations on the IR, modifying and refining it from one step to the next. We found that compiler flows for a machine such as the WSE challenges one of the underlying assumptions of MLIR. Work such as \cite{bisbas2024shared}, targeting, conventional architectures is able to leverage MLIR's rich GPUs and CPUs based infrastructure, and it is highly effective to entirely lower out of the representation of the problem, in this case stencils, at an early stage in the pipeline. By contrast, we found that for dataflow, cache-less, architectures such as the WSE it was important to retain high level information about the nature of the kernel, in this case stencils, throughout the majority of the pipeline in order to be able to use this to most effectively generate a representation that most suits the WSE.

This section presents a strategy by which a stencil IR representation is iteratively lowered by applying our set of transformations, leveraging both domain-specific and domain-agnostic intermediate dialects described in Section~\ref{sec:dialects}. Listing \ref{lst:mlir_stencil_b} sketches the stencil MLIR representation, generated from the Fortran code in Listing \ref{lst:fortran_stencil_example} by \cite{brown2023fortran}. At the initial stage of our compilation pipeline, the stencil dialect provides a mathematical description of the computation using value semantics, where variables represent independent copies of data without side effects or consideration of operational concerns such as how these are allocated to memory. A major contribution of our work is how this stencil representation is transformed by the compiler into a form that is optimized for the Cerebras WSE, and we use Listing \ref{lst:mlir_stencil_b} as a running example throughout this section to illustrate these transformations.

\begin{listing}[htb]
\begin{mlir*}{fontsize=\scriptsize}
  ^0(%
    "stencil.return"(%
}) : ... -> !stencil.temp<[0,254]x[0,254]x[0,510]xf32>
\end{mlir*}
\caption{Example stencil code used as a running example in this section.
\label{lst:mlir_stencil_b}}
\end{listing}

As per the green dashed lines in Figure \ref{fig:overall_flow}, there are five main groups of transformations that operate upon the IR to ultimately result in \emph{csl-ir}, which can be compiled for the WSE. We now explore each of these groups.

\subsection{Group 1: Decomposition and data dependencies}
The first group of transformations in our approach decomposes the three-dimensional representation of the stencil onto the WSE's two-dimensional grid of PEs and determines the data dependencies for stencil calculations on each PE. The overarching strategy is for a three-dimensional stencil of $x\times y\times z$ to be transformed into a two-dimensional stencil of size $x\times y$ such that each PE holds a column of $z$ values. This choice of decomposition has been demonstrated to be effective on the WSE \cite{jacquelin2022scalable}, and although mapping several columns of $z$ values to each PE is theoretically possible, this approach was adopted due to the large number of PEs and small local memory size (48 kB).

Decomposition at the IR level is driven by the existing pass \emph{distribute stencil} and we found that whilst this was initially intended for mapping stencils onto an HPC cluster via MPI \cite{bisbas2024shared}, the same abstract logic also applies to decomposition across the WSE's $x$ and $y$ sized grid. Ultimately, this results in the addition of the \texttt{\small DMP} dialect into our IR with \mlirinline{dmp.swap} operations inserted into the IR before each \mlirinline{stencil.apply} as shown in Listing~\ref{lst:mlir_stencil_c}. This operation indicates the necessary data exchanges that must occur before a specific \texttt{\small apply} can be performed, and as such describes remote data dependencies. The result of this transformation is a decomposition of the mathematical representation across the WSE, with high-level communication operations denoting when data must exchanged between PEs before a stencil operation can be performed.

\begin{listing}[htb]
\begin{mlir*}{fontsize=\scriptsize}
  #dmp.topo<254x254>, false>, swaps = [...]}
  (!stencil.temp<[-1,2]x[-1,2]xtensor<512xf32>>) ->
  !stencil.temp<[-1,2]x[-1,2]xtensor<512xf32>>
  ^0(%
    "stencil.return"(%
}) : ... -> !stencil.temp<[0,1]x[0,1]xtensor<512xf32>>
\end{mlir*}
\caption{After distributing $x$, $y$ dimensions across PEs and tensorizing $z$ dimension.
\label{lst:mlir_stencil_c}}
\end{listing}

The second transformation in this first group is to \emph{tensorize} over the $z$-dimension, transforming the three-dimensional grid of FP32 scalars into a two-dimensional grid of FP32 tensors. The MLIR \texttt{\small tensor} dialect provides a way of representing abstract collections of data. This transformation maintains value semantics of the data representation, but also provides rank-polymorphism because operations in the \texttt{\small arith} dialect work across tensors of values, rather than scalars. This rewrite is illustrated in Listing~\ref{lst:mlir_stencil_c}, where converting the stencil into a two-dimensional stencil operating over tensors means that each stencil element (a column of tensors) is mapped to an individual PE on the WSE.

\subsection{Group 2: Realize placement on and communication for WSE}

This group of transformations maps the decomposed mathematical representation of the problem, that results from the first group of transformations, onto a physical representation of the WSE and converts data dependencies into communications between PEs. As described in Section \ref{sec:async}, due to limited buffer sizes communications must occur in chunks and this also provides an opportunity to overlap communication and computation for performance. Consequently, the \emph{stencil to csl-stencil} transformation converts from the \texttt{\small stencil} to the \texttt{\small csl-stencil} dialect which involves all \mlirinline{dmp.swap} operations being replaced by \mlirinline{csl-stencil.prefetch}. These \mlirinline{csl-stencil.prefetch} operations are added into their corresponding \mlirinline{csl-stencil.apply} which itself combines both compute and communication.

The compute region of \mlirinline{stencil.apply} is split into two, the first region describing compute based on chunks of data that are held remotely and the second region on data held locally. This is illustrated in Listing~\ref{lst:mlir_stencil_e}, where communication involves two chunks, with the resulting \emph{acc} accumulator from the first region then consumed by the second region. There are two purposes to the first, remote data handling, region firstly the packing data, in the case of Listing~\ref{lst:mlir_stencil_e} inserting the received chunk of data into the result tensor, and secondly undertaking optional compute on received data before it is packed.

\begin{listing}[htb]
\begin{mlir*}{fontsize=\scriptsize}
  [#csl_stencil.exchange<to [1, 0]>], num_chunks = 2 : i64}> ({
  ^0(%
    ...
    csl_stencil.yield %
}, {
  ^1(%
    csl_stencil.yield %
}) : ...
\end{mlir*}
\caption{After transformation from the \texttt{\small stencil} to \texttt{\small csl-stencil} dialect.
\label{lst:mlir_stencil_e}}
\end{listing}

The other transformation in this group, \emph{wrap csl-stencil in csl-wrapper}, generates the layout metaprogram which maps kernels to PEs on the WSE, along with program-wide constant parameters extracted from the csl-stencil operations. Although \texttt{\small csl-wrapper} is a domain-agnostic dialect, this pass nevertheless populates it with highly domain-specific information from the \texttt{\small csl-stencil} dialect, which it wraps and retains.

\subsection{Group 3: Memory realization within a PE}\label{sec:cs-compiler-mem}

Until this point compute operations on the stencil have been expressed as value semantics where the focus is on the values themselves and operational details, such as where these are held in memory, is abstracted. This group of transformations converts the IR into reference semantics, which involves mapping tensors to physical memory allocation by way of transforming them to \texttt{\small memref}s. A major reason for doing this is that CSL mathematical operations follow the Destination-Passing Style (DPS) form, where they operate on physical memory, reading inputs from and storing results to physical memory that has been passed as operands. By contrast, \texttt{\small tensor}s are immutable, where arithmetic operations involving tensors always result in a new tensor.

A transformation pass rewrites our proposed dialects to reference semantics, undertaking \emph{partial bufferization}, before MLIR's bufferization is used to convert \texttt{\small tensor}s to \texttt{\small memref}s. If a computation requires additional intermediate buffers then MLIR's bufferization pass acts as a fail-safe that will automatically allocate these as required. Indeed, this demonstrates one of the benefits of using MLIR more generally, where extensive functionality is already provided by the MLIR ecosystem for common operations such as materialization, and this can be reused rather than having to reinvent it within an entirely bespoke compiler stack.

\begin{listing}[htb]
\begin{mlir*}{fontsize=\scriptsize}
  ^1(%
    linalg.add ins(%
    linalg.mul ins(%
    csl_stencil.yield %
}) : ...
\end{mlir*}
\caption{After transformations in the third group of dialects that realize memory within a PE.
\label{lst:mlir_stencil_d}}
\end{listing}

The \texttt{\small arith} dialect does not support CSL's DPS form of operations, whereas the linear algebra \texttt{\small linalg} dialect does. We therefore convert compute operations in the \texttt{\small arith} dialect to their counterparts in the \emph{linalg dialect}. Listing \ref{lst:mlir_stencil_d} illustrates the IR after these transformations have been run, where it can be seen that \texttt{\small memref}s replace \texttt{\small tensor}s and these are provided to \texttt{\small linalg} operations as operands. Furthermore, to make best use of the limited amount of memory on the WSE, the \emph{acc} \texttt{\small memref} is reused in Listing \ref{lst:mlir_stencil_d} as input, to hold intermediate results, and final calculation results.

\subsection{Group 4: Map to the actor execution model}

Figure \ref{fig:time_step_loop} illustrated that in order to communicate within a CSL program one must structure their code as tasks that are called when corresponding messages arrive. It was highlighted in Section \ref{sec:actor_exec_model} that it is effective to view the WSE as following the actor model, where PEs are hardware actors and tasks software actors. This group of transformations lowers to the WSE's asynchronous execution model by splitting up \mlirinline{csl-stencil.apply} operations into their constituent activities and mapping these to actors. Because \mlirinline{csl-stencil.apply} operations perform asynchronous communication on the WSE, their remote-data and local-data subregions are lowered into software actors which are activated, respectively, each time chunks of remote data are received, and once at the end when all chunks from all neighbors have been received. This is very similar to the \emph{stencil.communicate} function called by \emph{seq\_kernel0} and \emph{seq\_kernel1} in Figure~\ref{fig:time_step_loop}.

The second actor resulting from this lowering to actors typically processes data held locally and is executed exactly once as the last part of the operation. Similarly to Figure~\ref{fig:time_step_loop}, it is from this second task that control flow proceeds, where the successive function is invoked by this software actor.

As described in Section \ref{sec:async}, to enable naturally synchronous stencils to operate in an asynchronous fashion one is unable to have a top-level timestep loop, or even successive \mlirinline{csl-stencil.apply} operations without a loop. This is because the \mlirinline{csl-stencil.apply} operation triggers asynchronous communications after which program flow does not continue in a synchronous manner, and both these configurations are required by the benchmarks explored in Section \ref{sec:evaluation}. The second transformation of this group converts the top-level control flow enclosing \mlirinline{csl-stencil.apply} operations into a control-flow task graph, comprising software actors which consist of callable zero-parameter functions as well as local tasks in our \texttt{\small csl-ir} dialect. For instance, a top-level timestep \texttt{\small scf.for} loop is lowered into a control-flow graph consisting of CSL functions instead of basic blocks.

\subsection{Group 5: Lowering to the csl-ir dialect}
\label{sec:lowering_to_cslir}

The last group of transformations will lower to the \texttt{\small csl-ir} dialect from which CSL code is printed out. The first transformation of this fifth group, \emph{linalg to csl-ir}, lowers compute operations, which are in the \texttt{\small linalg} dialect, into corresponding operations in the \texttt{\small csl-ir} dialect. These operations are expressed across arrays of data, and the naive approach would be to generate a loop that performs an arithmetic operation for each element. However, best practice on the WSE is to instead use CSL's Data Structure Descriptors (DSDs), which are essentially affine iterators over buffers via native hardware support. CSL provides a set of high-throughput DSD arithmetic builtin functions that deliver a significant performance improvement compared to computations that are not based on DSD builtins. For instance, via DSDs, the multiply and add operations shown in Listing~\ref{lst:mlir_stencil_d} are lowered to \texttt{\small @fmacs(dest, src1, src2, f32\_val)}, providing an FP32 fused multiply-accumulate.

At this point, the stencils and compute components have been transformed into the \texttt{\small csl-ir} dialect. However, memory is still represented by the \texttt{\small memref} dialect. While \texttt{\small memref} types are used throughout the \texttt{\small csl-ir} dialect, operations on \texttt{\small memref}s, such as allocation and deallocation, need to be lowered to the \texttt{\small csl-ir} dialect. Furthermore, DSDs are initialized as views and subviews onto buffers. Consequently, the \emph{lower memref to dsd} pass generates DSD definitions on top of memref buffer allocations that are used throughout inbuilt compute functions as well as for communication.

The last step of this group lowers the top level \linebreak\mlirinline{csl-wrapper.module} operation to \emph{csl-ir}. This transformation generates two output modules, the first comprises \emph{csl-ir} operations for the layout metaprogram from its abstract representation, and the second is the \emph{csl-ir} operations generated in preceding passes.

\subsection{Runtime communications library}
\label{sec:comms_library}

Data exchanges between hardware actors (PEs) are handled by a CSL-based library that we designed using the partitionable communication strategy developed by \citeauthor{jacquelin2022scalable} \cite{jacquelin2022scalable} for star-shaped stencils of up to three dimensions at variable stencil sizes. The library encapsulates boiler-plate code for sending and receiving data split into chunks of configurable sizes. Initially, it schedules asynchronous send and receive data exchanges in all four directions, using multiple internal \emph{tasks} per direction to handle the completion of asynchronous communication steps and updating of routing patterns, and which in turn trigger user-provided callbacks. CSL's staged compilation mechanism provides sufficient flexibility and power to compile arbitrary star-shaped stencil patterns, allowing us to focus on code generation on the software-actor level, internal to each PE. Supporting different stencil shapes (for instance, box-shaped communication patterns) would require a library update. Alternatively, in work complementary to ours \citeauthor{sai2024automated}~\cite{sai2024automated} present software-based routing patterns, whereas code generation for hardware routing configurations and switching updates remains future work, even in the confined case of stencil patterns.

\subsection{Optimization passes}\label{sec:optimization}

Whilst much of this section has focused on transformations to recast the mathematical stencil representation to a form that suits execution on the WSE, several additional transformations were introduced solely for the purpose of optimization at the stencil and arithmetic levels. The first of these, \linebreak\texttt{\small stencil-inlining}, is executed at the start of the compilation pipeline and attempts to merge consecutive \mlirinline{stencil.apply}s into a single fused kernel. This is an existing pass that was developed in \cite{bisbas2024shared} to remove the unnecessary overhead of host-device context switching between stencils that are consecutive. The transformation ensures all outputs of the first stencil are passed through to the second, even if values are unaffected, and the operations performed by both stencils are then merged into a single \mlirinline{stencil.apply}. Being able to leverage this existing transformation is an example of one of the key benefits of integrating with a wider compiler ecosystem such as MLIR, as it avoids duplication of effort developing common passes. In the case of the UVKBE benchmark, that we explore in Section \ref{sec:evaluation}, this pass is able to merge all \mlirinline{stencil.apply}s into one single operation. The fused kernel is subsequently split later in the pipeline according to buffer communications, and to allow for potential interleaving of communication and computation.

\texttt{\small Varith} is an xDSL-specific dialect which represents multiple arithmetic additions or multiplications as a single variadic arithmetic operation. We employ this dialect early in the compilation pipeline to represent arithmetic operations where possible because it significantly reduces the complexity associated with manipulating them, e.g. when splitting the computation into regions for separate processing of locally and remotely held data. Furthermore, the \texttt{\small varith} dialect is required by the \texttt{\small varith-fuse-repeated-operands} transformation which we leverage to convert a series of consecutive additions of the same argument into a multiplication. This optimization is especially important for codes such as the Acoustic kernel that we explore in Section \ref{sec:evaluation}, as it can eliminate multiple arithmetic operations, in the case of that kernel we were able to replace three DSD addition operations with a single multiplication.

Moreover, the \texttt{\small varith-fuse-repeated-operands} transformation also enables us to fuse multiplications and additions into an \emph{fmac} operation. When this transformation is used in conjunction with \texttt{\small linalg} dialect to model DSD operations, it makes it possible for our \texttt{\small linalg-fuse-multiply-add} transformation to identify multiplication and addition operations that can be combined. Ultimately, the result of applying this transformation is that the \texttt{\small @fmacs} CSL intrinsic, that was mentioned in Section \ref{sec:cs-compiler-mem}, is generated for suitable chains of arithmetic operations. Due to the prevalence of a multiplication followed by addition in stencils, we have found that this optimization results in a large number of multiplication and addition operations being converted into \texttt{\small @fmacs} which delivers improved performance.

An efficient lower-level optimization strategy is to check whether the same reduction function is applied across the entire stencil shape. In this case, we can generate highly optimized code using tensor broadcasting techniques on the accumulator buffer, expanding it with virtual dimensions that match the communication buffer, so as to perform a one-shot reduction of the entire communication buffer in a single builtin function call. Conversely, if points in the stencil shape needs to be processed using different functions, or if a subset of points is altogether ignored, the compiler instead needs to obtain separate pointers into the communication buffer and generate individual builtin function calls to process communicated data, which is less performant. Another low-level optimization is to promote coefficient application into the communication code where possible, to interleave communication and computation. Effectively, instead of moving incoming data from an input queue to PE memory with a \texttt{\small @mov}-like builtin, we can instead multiply incoming data from the input queue to memory, by a set of coefficients. These two low-level techniques can be used in combination.

\section{Evaluation}\label{sec:evaluation}

In this section we compare the performance of our approach on both a Cerebras CS-2 and CS-3 system against CPU and GPU baselines. For each benchmark, we considered three different problem sizes, classified by the number of elements in the X and
Y dimensions as: small -- $100\times100$, medium -- $500\times500$ and large
-- $750\times994$. The large size was chosen to fully occupy the PE grid on the
WSE2. The Z dimension varies for each of the five following benchmarks:

\begin{description}

  \item[Jacobian \emph{(Flang)}] is extracted from Fortran code solving
      Laplace's equation for diffusion in 3D. This is a six-point stencil, and is run for 100,000 iterations with 900 elements in the Z dimension.

  \item[Diffusion \emph{(Devito)}] solves the heat diffusion equation. Expressed in Devito's Python-based DSL, this is a 3D 13-point stencil running for 512 iterations with 704 elements in the Z dimension.

  \item[Acoustic \emph{(Devito)}] models the isotropic acoustic wave equation in Devito using a 2\textsuperscript{nd} order approximation in time. Similar to `Diffusion' this is a 3D 13-point stencil, run for 512 iterations with 604 elements in the Z dimension.

  \item[25-Point Seismic \emph{(Cerebras)}] is a seismic kernel written in CSL using a 1\textsuperscript{st} order approximation in time. Translated from \citeauthor{jacquelin2022scalable}, this kernel provides a direct comparison against manually tuned CSL. This is a 3D 25-point stencil, ran for 100,000 iterations with 450 elements in the Z dimension.

  \item[UVKBE \emph{(PSyclone)}] includes four fields, two of which need to be communicated across PEs as well as two consecutive \mlirinline{stencil.apply} operations. This benchmark is written in Fortran via the PSyclone DSL and is run for a single iteration with 600 elements in the Z dimension.

\end{description}

\begin{figure}
  \centering
  {\includegraphics[width=\linewidth]{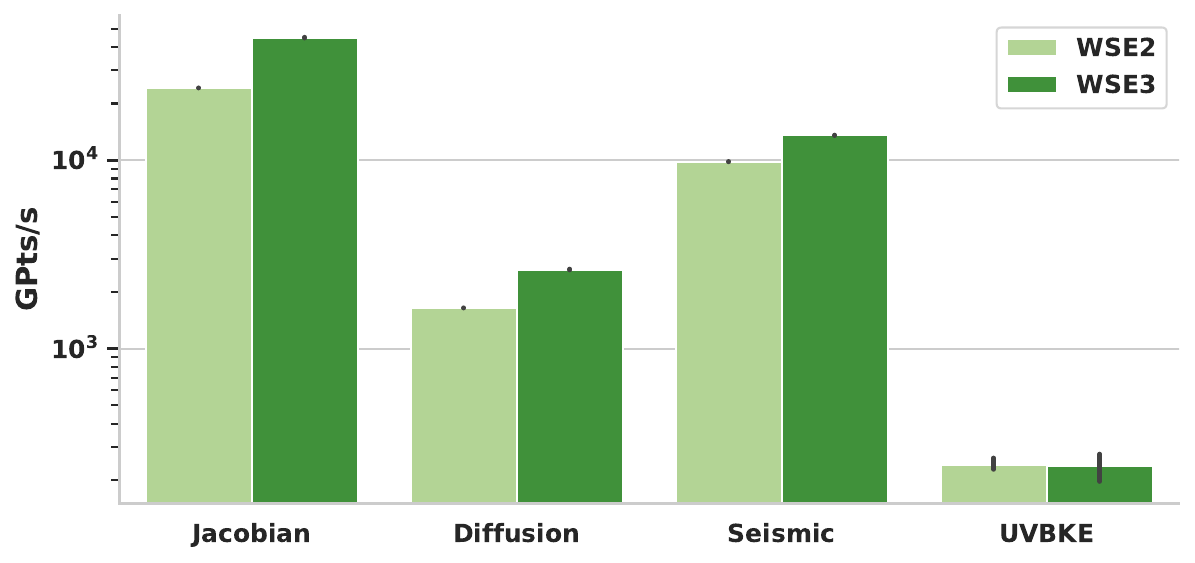}}
  \caption{Performance comparison of WSE2 and WSE3 across benchmarks running the large problem size using our approach. Our code generator exploits the latest hardware features for optimal performance.}
  \label{fig:eval_hardware}
\end{figure}

For all benchmarks, we use 32-bit single-precision floating point. Reported results are the average throughput measured in GPts/s (a.k.a GCells/s) over three runs. It should be highlighted that Figures \ref{fig:eval_hardware}, \ref{fig:eval_us_vs_them} and \ref{fig:cs_3-vs-others} include error bars (CI 95\%) but these are very small because there were no significant outliers between runs. We use xDSL version v0.35 and target Cerebras SDK 1.3.0 and Cerebras ML Software 2.4. Figure \ref{fig:eval_hardware} reports performance across four of these benchmarks with the large problem size from Flang (Jacobian), Devito (Diffusion), Cerebras (Seismic), and PSyclone (UVKBE) using our compilation approach on the WSE2 and WSE3. Benchmarks are run for 100000, 512, 100000 and 1 iteration(s) respectively. The major reason for the performance difference between the WSE2 and WSE3 is the utilization of the WSE3's upgraded switching logic. Hardware limitations on the WSE2 in the switch configuration require each PE to transmit data to itself as well as to its neighbors (on each of the east, west, north, and south routes) \cite{jacquelin2022scalable}. On the WSE3 this is no longer necessary, and our WSE3 communications library exploits this architectural improvement.

\subsection{Our approach vs hand-written WSE code}

\begin{figure}
  \centering
  {\includegraphics[width=\linewidth]{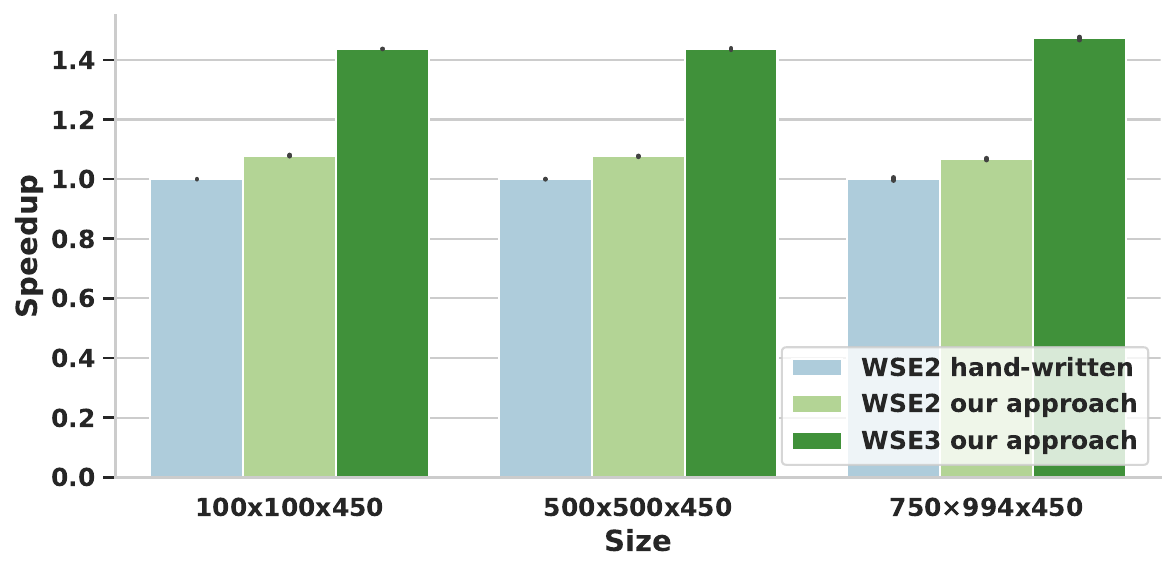}}
    \caption{Our compiler-based approach for the seismic benchmark plotted against the hand-written equivalent developed by Cerebras \cite{jacquelinCslexamplesBenchmarks25ptstencil}. Efficient communication and task utilization gives our generated code the performance edge.}
  \label{fig:eval_us_vs_them}
\end{figure}

In collaboration with Cerebras engineers, \citeauthor{jacquelin2022scalable}~\cite{jacquelin2022scalable} developed a hand-written, optimized seismic
modeling benchmark based on a 25-point stencil. While available
in the Cerebras SDK for the WSE2, it is not supported on the WSE3~\cite{jacquelinCslexamplesBenchmarks25ptstencil}. \autoref{fig:eval_us_vs_them} reports
performance of this hand-written kernel on the WSE2 for different problem sizes
in comparison to an equivalent implementation generated by our approach for the
WSE2 and WSE3. \cite{jacquelin2022scalable} reported that the hand-written
implementation achieved 28.2\% of theoretical peak performance on the WSE2, and
the version generated by our compilation pipeline outperforms this by up to
7.9\%. The performance is affected by several factors. Our generated version is
more memory efficient, allowing communication in a single chunk where the
hand-written version uses two. However, data from
\citeauthor{jacquelin2022scalable} suggests that this may not account for all
performance differences. Our compilation approach only communicates data that is
required by the calculation. For this benchmark, this means that the first and
last column values are not communicated, whereas the hand-written code transmits
the full column of values, including those which are not required. Our
communications library also manages tasks more efficiently, reducing the total number
of tasks required by approximately 50\%, which reduces task activation overhead.

The hand-written code is restricted to the WSE2 due to architectural hardware generations that require kernel rewriting. By contrast, our approach seamlessly supports both the WSE2 and WSE3. Overall, the code generated for the WSE3 outperforms the WSE2 implementation by up to 38.1\%, primarily due to the architectural improvements described previously. As described in Section \ref{sec:related}, \cite{sai2024automated} proposed a Python-based stencil DSL for the WSE leveraging an entirely bespoke compilation pipeline. Their compiler is not publicly available, but in \cite{sai2024automated} they compare against the same manually implemented 25-point stencil kernel as \autoref{fig:eval_us_vs_them}. They report marginally worse performance than the manually implemented code, and the argument made in \cite{sai2024automated} is around programmer productivity. By contrast, as demonstrated in this section, for this same kernel our approach delivers both programmer productivity and performance advantages.

\subsection{Performance on WSE3 against GPUs and CPUs}

\begin{figure}
  \centering
  {\includegraphics[width=\linewidth]{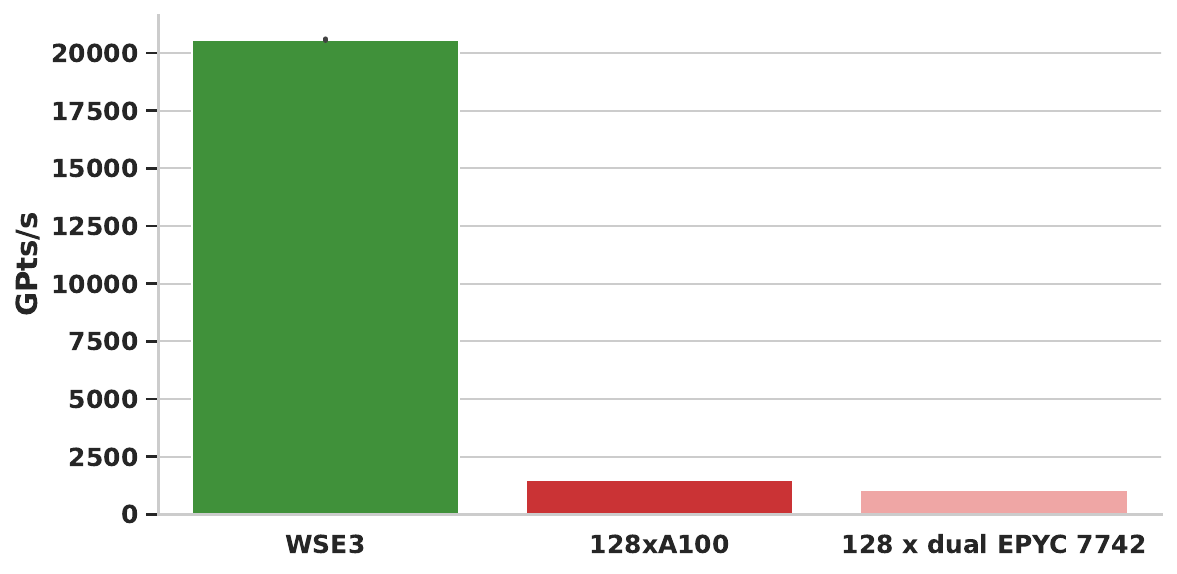}}
  \caption{Performance of Devito acoustic benchmark on the WSE3 (large problem size, 100,000 iterations) compared to MPI + OpenACC kernels on 128 A100 GPUs on Tursa and MPI + OpenMP kernels on 128 CPU nodes on the ARCHER2 Cray-EX, from \cite{bisbas2025automated}. A single wafer delivers throughput as a large GPU or CPU cluster, showcasing high throughput capabilities of the WSE3.}
  \label{fig:cs_3-vs-others}
\end{figure}

To provide a comparison of our approach on the WSE3 against traditional HPC architectures, GPUs and CPUs, we evaluated the performance of Devito's \emph{acoustic} benchmark against results reported by \cite{bisbas2025automated}.
Their study evaluated MPI + OpenACC kernels on 128 Nvidia A100 GPUs on the Tursa supercomputer, and MPI + OpenMP kernels on 128 CPU nodes of the ARCHER2 CPU based Cray-EX supercomputer. Each Tursa node contains two 24-core, AMD 7413 EPYC processors and four NVIDIA Ampere A100-80 GPUs connected via NVLink, and four 200 Gbps Infiniband links connecting the nodes. Each ARCHER2 node contains two 64-core, AMD Zen2 (Rome) EPYC 7742 processors, nodes, connected via HPE's proprietary Slingshot interconnect. Experimental setup used version 23.5-0 of Nvidia's nvc++ compiler on Tursa, and Cray's Clang version 11.0.4 on ARCHER2. All experiments used single-precision floating-point arithmetic.

\citeauthor{bisbas2025automated} used significantly larger grid sizes, $1024^3$ and $1158^3$ points on the CPU and GPU respectively, giving them a slight advantage by lowering communication overhead. Figure \ref{fig:cs_3-vs-others} demonstrates that the WSE3 achieves a significantly faster throughput for a specific problem size, leading the race for a faster solution against 128 A100s and 128 CPU nodes (16384 CPU cores). The WSE3 is around 14 times faster to solution than 128 A100s, and 20 times against 128 CPU nodes.

We acknowledge that this is not a strictly direct comparison due to the differing problem sizes. In the strong scaling study by \cite{bisbas2025automated}, decomposing the problem across GPUs results in each device computing a smaller sub-domain with a higher communication-to-computation ratio, thereby preventing full exploitation of single-GPU potential. By contrast, the WSE3 processes the entire domain on a single wafer. While a weak-scaling comparison—where each GPU solves a larger problem—would likely reduce the WSE3's speedup, the advantage would remain significant. Furthermore, the GPU baseline relied on OpenACC rather than hand-tuned CUDA (available only in proprietary versions of Devito), thus, not exercising its full potential. Nevertheless, this experiment is yet another demonstration that in the race for "time-to-solution for a specific problem size", the WSE3 outperforms large clusters of traditional hardware \cite{tramm2024efficient, santos2024breaking, perez2025breaking}.

Indeed, if one were to assume perfect scaling on the CPU then around 50\% of the ARCHER2 supercomputer would be required to match performance of a single WSE3 for this benchmark. Furthermore, whilst these performance numbers are impressive, it is also important to highlight that the Devito code here is running unchanged across the WSE3, GPUs and CPUs.

\subsection{WSE3 roofline performance}
\label{sec:roofline}
A roofline plot visualizes a kernel's performance in FLOPs/s against its arithmetic intensity in FLOPs/byte, bounded by the system's peak compute and memory bandwidth limits. The dashed lines represent the theoretical maximum of the machine, and applications that fall under the initial diagonal part of this line are bound by memory performance, and those under the horizontal part are limited by compute. The higher the performance of a benchmark then the closer it is to the theoretical maximum. Indeed, it was a roofline plot that \citet{ltaief2023scaling} used to argue the memory benefits of the WSE, as they observed codes moved from under the diagonal portion, memory bound, on other architectures to the horizontal part, compute bound, on the WSE.

\begin{figure}
  \centering
  {\includegraphics[width=\linewidth]{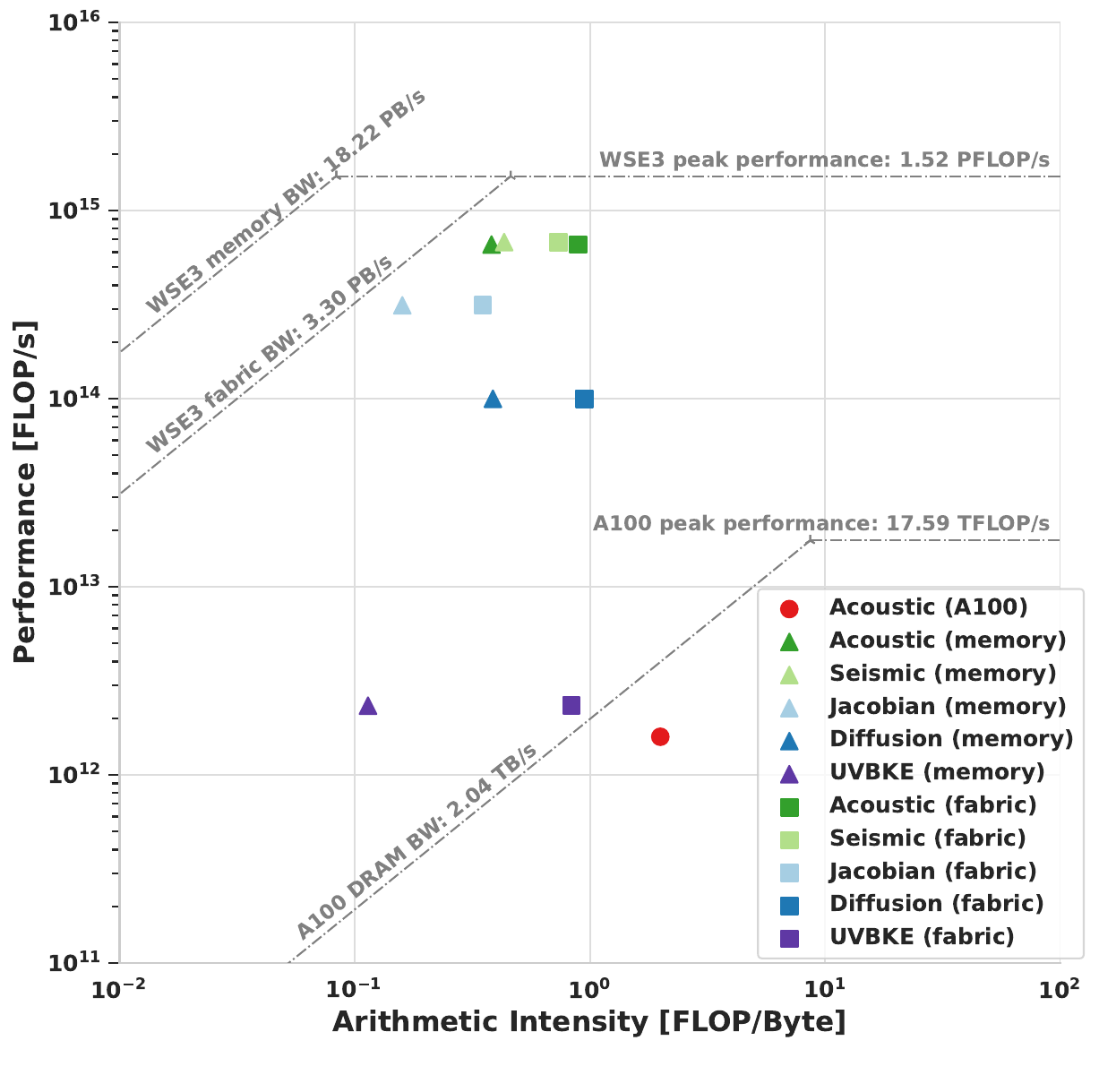}}
    \caption{Roofline plot of our five benchmarks on the WSE3 and the acoustic benchmark on a single A100.}
  \label{fig:roofline}
\end{figure}

Figure \ref{fig:roofline} plots the roofline of the WSE3 with our five benchmarks and the acoustic benchmark on a single A100 GPU. On the WSE3 local memory has a higher bandwidth than the communication fabric, and-so for each WSE benchmark we plot two values; the first based on all data accesses being from local memory and the second all data accesses being via fabric. It can be seen that for all WSE3 benchmarks that assume data resides in memory then the benchmarks are compute bound, and all apart from the Jacobian are also compute bound when assuming all data accesses are via the fabric. By contrast, we also plot the acoustic benchmark on the A100 GPU from \cite{bisbas2025automated} and it can be seen that this code is memory bound on the A100. Effectively, the approach described in this paper has enabled the user to move their kernel to an architecture which delivers much better performance, a major reason being that the code is no longer memory bound, without requiring any application-level changes. \citeauthor{jacquelin2022scalable}~\cite{jacquelin2022scalable} observed compute-bound performance for a 25-point seismic stencil, and our experiments show that the same finding applies to a larger variety of stencil programs.

\subsection{Developer productivity}

As described in Section \ref{sec:async}, writing code for the WSE is difficult and adds significant complexity not least because the programmer must rethink their execution model. Consequently, from a programmer productivity perspective, it can take significant time to develop and optimize code for the architecture \cite{brown2022exploring}. However whilst it does not provide a full picture, a lines of code, LoC, metric provides some indication as to the effort required.

\begin{table}[htb]
    \begin{center}
    \begin{tabular}{|c|c|c|c|}
    \hline
       \textbf{Benchmark} & \makecell{\textbf{CSL kernel} \\ \textbf{only} \\ \textbf{\textit{(LoC)}}} & \makecell{\textbf{CSL} \\ \textbf{entire} \\ \textbf{\textit{(LoC)}}} & \makecell{\textbf{DSL \& our} \\ \textbf{approach} \\ \textbf{\textit{(LoC)}}} \\
      \hline
    25-point Seismic & 196 & 980 & 81 \\
    Acoustic & 211 & 995 & 81 \\
    Diffusion & 192 & 976 & 40 \\
    Jacobian & 180 & 964 & 28 \\
    UVKBE & 203 & 987 & 44 \\
    \hline
    \end{tabular}
    \caption{Lines of Code (LoC) comparison between generated CSL and programming using a DSL with our compilation approach.}
    \label{tab:loc-comparison}
    \end{center}
\end{table}

Table \ref{tab:loc-comparison} reports a lines of code comparison for each benchmark between the generated CSL and that required when writing code in a DSL and using our compilation approach. We provide two numbers for the CSL code, \emph{CSL kernel only} comprises the kernel alone without any support for placement on the WSE, communications or interaction with the host. \emph{CSL entire} is the full CSL code length containing both the kernel and all the required support functionality in our runtime library that was described in Section \ref{sec:comms_library}. It can be seen that, regardless of the benchmark and DSL, using our approach requires significantly reduces the amount of code that must be written. Furthermore, it should be highlighted that the DSL source code would likely be much more accessible for a scientific programmer to understand, whereas not only is CSL very architecture specific but it also tends to be difficult to differentiate the scientific stencil code from all the WSE specific implementation details that surround it.

\section{Related Work}
\label{sec:related}

\citeauthor{rocki2020fast} \cite{rocki2020fast} explored stencil codes on the first iteration of the WSE, the CS-1.
The authors ported the main components of the iterative BiCGSTAB Krylov subspace method, the SpMV, AXPY and inner product, to the WSE. This work was before Cerebras released CSL and so was undertaken using a proprietary programming interface, achieving 0.86 PFLOPS, over 200 times the performance of the NETL Joule supercomputer. Recognizing the programmability challenge of the WSE, \citeauthor{woo2022disruptive}~\cite{woo2022disruptive} followed up this early work by developing a Python-based API for programming the WSE. Crucially, their API maintained the performance of the CS-2 by delivering over two orders of magnitude runtime reduction on the WSE2 compared to the NETL Joule 2.0 supercomputer. Our work addresses two major limitations of \citet{woo2022disruptive}, firstly in their approach programmers must convert their code to use a specific API, whereas our work places the emphasis on the DSL developer to integrate with the \texttt{\small stencil} dialect and then, without any user code changes, these will run on the WSE. The second major difference is that \citet{woo2022disruptive} generated assembly level code, which itself is under NDA, rather than CSL. Instead, by targeting CSL, our approach is portable between generations, such as WSE2 to WSE3, and we can also publish the source code of the compiler, as it relies only on public APIs.

A Python-based stencil DSL for the WSE was proposed by \citeauthor{sai2024automated}~\cite{sai2024automated} which, similarly to our approach, produces CSL code. Their focus is largely on generating a software-based routing algorithm for different stencil shapes in a manner complementary to our work, whereas the generation of compute or overarching program structure is not discussed. Similarly to \citet{woo2022disruptive}, they developed their own entirely bespoke compiler whereas our approach provides a series of building blocks within the MLIR ecosystem. Furthermore, unlike previous work (see \cite{sai2024automated, woo2022disruptive}) we decouple the frontend programming technology from the compilation flow targeting the WSE, therefore application code can remain unchanged rather than first needing to port into a new DSL.

SPADA \cite{gianinazzi2025spada} is a spatial dataflow language targeted at the WSE, specifically the CS-2, where \citeauthor{gianinazzi2025spada} developed a bespoke compiler and integrated the GT4Py stencil DSL atop this. SPADA and our approach are complimentary, where SPADA heavily optimizes the lower levels of the stack by providing a WSE specific abstraction and optimizations. Conversely, by leveraging MLIR we progressively lower through multiple abstraction levels towards CSL, exploiting each of these levels and existing compiler building blocks where applicable to generate the most appropriate target code for both CS-2 and CS-3 versions of the WSE architecture. Fundamentally, SPADA aims to solve a different problem to our work, where we prioritize sharing of infrastructure between frontends to promote ease of integration. Consequently, there is a greater semantic gap in \cite{gianinazzi2025spada} between the GT4Py stencil DSL and SPADA, requiring more work in each of the frontends to connect stencils to SPADA and risking duplication. By contrast using our approach once the stencil dialect has been generated then our compilation pipeline handles the entirety of the transformation from this to the WSE in a frontend agnostic manner. Indeed, it would be interesting in future work to integrate our work with SPADA, if they were to develop an MLIR dialect, as this would enable stencil DSL frontends to transparently exploit the WSE specific abstractions and optimizations developed by \citeauthor{gianinazzi2025spada}.

Whilst differences in experimental setup make performance comparisons somewhat difficult, given the complimentary nature of SPADA and our work it is still interesting to highlight key aspects. Using our approach, as described in Section \ref{sec:roofline}, a 3D Jacobian which implements LaPlace's equation for diffusion delivers 169 TFLOP/s on the CS-2 and 313 TFLOP/s on CS-3, compared to a 2D Laplacian in \cite{gianinazzi2025spada} delivering 120 TFlop/s on the CS-2. UVKBE is the highest performing kernel in \cite{gianinazzi2025spada} delivering around 150 TFlop/s on the CS-2. This is greater performance than the UVKBE benchmark in our work, but we adopted the UVKBE benchmark from \cite{gysi2021domain} which only executes one iteration. This is different from UVKBE in \cite{gianinazzi2025spada} which has more opportunity to ameliorate overhead. Conversely, our highest performing kernel (25-point stencil) delivers 491 TFlop/s on the CS-2 and 678 TFlop/s on the CS-3. Irrespective, \citeauthor{gianinazzi2025spada} also demonstrate the significant programmer productivity benefits of driving the WSE from a stencil based DSL and deliver impressive reductions in the lines of code required compared to manually written CSL.

\citeauthor{jacquelin2022scalable}~\cite{jacquelin2022scalable} developed an optimized implementation of a 25-point stencil for seismic modeling on the CS-2, and \citeauthor{sai2023massively}~\cite{sai2023massively} implemented a 3D 9-point stencil, demonstrating how data can be indirectly transferred to PEs at diagonal offsets to support diagonal data dependencies for stencils of rank $1$. Furthermore, \citeauthor{sai2024matrix}~\cite{sai2024matrix} present a finite volume kernel that performs a three-dimensional 7-point stencil operation followed by a whole-fabric all-reduce operation. A cornerstone of our approach is to fold the techniques and best practices in these papers into the compiler, enabling non-expert programmers to benefit from such insights without requiring expertize in programming the WSE.

The scientific computing community has long explored abstraction layers to improve productivity, spanning high-level domain-specific approaches and low-loop/expression-level optimizations. Higher-level symbolic frameworks include the Python-based OpenSBLI \citep{jacobs2017opensbli} which employs Einstein notation for symbolic modeling, automating parallelism via the C/C++-based OPS library \citep{ops2018}. ExaStencils \citep{exaslangMPIKucuk:2016, exastencils_2020} combines layered symbolic DSLs for finite differences, supporting CPU/GPU architectures through code generation. STELLA \citep{Stella2015}, and subsequently GridTools \citep{gridtools2021}, uses mathematical discretization syntax notably enabling GPU acceleration for the COSMO weather model \citep{Thaler2019}.

However, none of these existing stencil frameworks support architectures such as the WSE, and in this work we demonstrate one of the major benefits of MLIR, where a range of frontend languages and DSLs can benefit from building blocks which lower an existing \texttt{\small stencil} dialect to the WSE. Indeed, stencil DSLs such as those above could integrate with our approach. Furthermore, whilst our approach is the first MLIR pipeline for the Cerebras WSE, there are several MLIR dialects and transformations developed for other novel architectures such AMD's AIEs~\cite{AMD-AIE}, Tenstorrent Tensix~\cite{TT-MLIR}, and \citet{wang2025mlir} provide MLIR infrastructure for CGRA compilers. However, whilst these also provide MLIR building blocks for these different architectures, unlike our approach on the Cerebras WSE, none of them support acceleration of unmodified code in existing languages and DSLs on the target accelerator.

\section{Conclusions and further work}

In this paper we have presented an MLIR lowering pipeline for stencil codes, demonstrating how one can effectively target the Cerebras WSE in code generation based upon existing high-level stencil domain specific abstractions without code level changes. We evaluate using five benchmarks from three different programming frontends, and demonstrated that our approach is competitive to, and can out-perform, a stencil hand-tuned, in part, by Cerebras engineers, and in comparison to a cluster of 128 Nvidia A100 GPUs and 128 nodes of the CPU-based ARCHER2 Cray-EX supercomputer managed to out-perform these by 14x and 20x, respectively.

Prior to this work considerable effort has always been required to port a code to WSE3, not just requiring a rewrite of the code in CSL but furthermore a fundamental recasting of the algorithm to suit the asynchronous, actor-based nature of the Cerebras system. During this work we made the observation that adopting the actor-based paradigm helps provide an execution model of communication and computation on the WSE that enables consistency. In future work, we consider this important to be explored further by the community, not just as the potential programming model for the WSE but moreover how it might characterize some of the problems encountered when targeting a range of other CGRA-like accelerators. Our implementation is available as open-source under \url{https://github.com/xdslproject/wse-stencil} which uses v0.35 of xDSL that contains many of the dialects and transformations discussed in this paper \url{https://github.com/xdslproject/xdsl}.

\begin{acks}
This work was supported by the \grantsponsor{EPSRC}{Engineering and Physical Sciences Research Council (EPSRC)}{https://www.ukri.org/councils/epsrc/} grant \grantnum{EPSRC}{EP/W007940/1} and a Royal Society of Edinburgh personal research fellowship award number 3271. We thank the Edinburgh International Data Facility (EIDF) at the University of Edinburgh for access to the CS-3 and Leibniz Supercomputing Centre (LRZ) for the CS-2. The authors would like to thank Cerebras for providing technical support, along with P. Kelly, E. Bauer and N. Heinimann for their thoughtful early feedback, and finally the shepherd and reviewers for their comments and support improving this work.
\end{acks}

\bibliography{references}

\end{document}

%% file: tex/setup.tex
\makeatletter
\patchcmd{\@addmarginpar}{\ifodd\c@page}{\ifodd\c@page\@tempcnta\m@ne}{}{}
\makeatother
\ifx\paperversion\paperversiondraft
  \makeatletter
  \if@ACM@journal
    \paperwidth=\dimexpr \paperwidth + 3.5cm\relax
    \oddsidemargin=\dimexpr\oddsidemargin + 0cm\relax
    \evensidemargin=\dimexpr\evensidemargin + 0cm\relax
    \marginparwidth=\dimexpr \marginparwidth + 3cm\relax
    \makeatletter
    \long\def\@mn@@@marginnote[#1]#2[#3]{%
      \begingroup
        \ifmmode\mn@strut\let\@tempa\mn@vadjust\else
          \if@inlabel\leavevmode\fi
          \ifhmode\mn@strut\let\@tempa\mn@vadjust\else\let\@tempa\mn@vlap\fi
        \fi
        \@tempa{%
          \vbox to\z@{%
            \vss
            \@mn@margintest
            \if@reversemargin\if@tempswa
                \@tempswafalse
              \else
                \@tempswatrue
            \fi\fi
              \rlap{%
                \if@mn@verbose
                  \PackageInfo{marginnote}{xpos seems to be \@mn@currxpos}%
                \fi
                \begingroup
                  \ifx\@mn@currxpos\relax\else\ifx\@mn@currxpos\@empty\else
                      \kern-\dimexpr\@mn@currxpos\relax
                  \fi\fi
                  \ifx\@mn@currpage\relax
                    \let\@mn@currpage\@ne
                  \fi
                  \if@twoside\ifodd\@mn@currpage\relax
                      \kern\oddsidemargin
                    \else
                      \kern\evensidemargin
                    \fi
                  \else
                    \kern\oddsidemargin
                  \fi
                  \kern 1in
                \endgroup
                \kern\marginnotetextwidth\kern\marginparsep
                \vbox to\z@{\kern\marginnotevadjust\kern #3
                  \vbox to\z@{%
                    \hsize\marginparwidth
                    \linewidth\hsize
                    \kern-\parskip
                    \marginfont\raggedrightmarginnote\strut\hspace{\z@}%
                    \ignorespaces#2\endgraf
                    \vss}%
                  \vss}%
              }%
          }%
        }%
      \endgroup
    }
    \makeatother
  \else
    \paperwidth=\dimexpr \paperwidth + 6cm\relax
    \oddsidemargin=\dimexpr\oddsidemargin + 3cm\relax
    \evensidemargin=\dimexpr\evensidemargin + 3cm\relax
    \marginparwidth=\dimexpr \marginparwidth + 3cm\relax
  \fi
  \makeatother
\fi

\usepackage[textsize=tiny,disable]{todonotes}
\usepackage[normalem]{ulem}

\makeatletter
\font\uwavefont=lasyb10 scaled 652
\newcommand\colorwave[1][blue]{\bgroup\markoverwith{\lower3\p@\hbox{\uwavefont\textcolor{#1}{\char58}}}\ULon}

\newcommand\highlight[2]{{\color{#1}{\colorwave[#1]{#2}}}}
\makeatother

\makeatletter
\newcommand\InFloat[2]{\ifnum\@floatpenalty<0\relax#1\else#2\fi}
\makeatother

\newboolean{inComment}
\setboolean{inComment}{false}

\ifx\paperversion\paperversiondraft
\newcommand\createtodoauthor[2]{
  \def\tmpdefault{emptystring}
  \expandafter\newcommand\csname #1\endcsname[2][\tmpdefault]{
    \ifthenelse{\boolean{inComment}}{
      \PackageError{paper-template}{Comments in comments not supported}{}
    }{}\setboolean{inComment}{true}
    \def\tmp{##1}
    \InFloat{
        \smash{
	  \marginnote{
	    \todo[inline,linecolor=#2,backgroundcolor=#2,bordercolor=#2]
	      {\textbf{#1 (Figure):} ##2}
          }
        }
    }{\ifthenelse{\equal{\tmp}{\tmpdefault}} 
      {\todo[linecolor=#2,backgroundcolor=#2,bordercolor=#2]{\textbf{#1:} ##2}\ignorespaces}
      {\ifthenelse{\equal{##2}{}} 
        {\highlight{#2}{##1}}
        {\highlight{#2}{##1}\todo[linecolor=#2,backgroundcolor=#2,bordercolor=#2]
	  {\textbf{#1:} ##2}
	}
      }
    }
    \setboolean{inComment}{false}
  }
}
\else
\newcommand\createtodoauthor[2]{%
\expandafter\newcommand\csname #1\endcsname[2][]{##1}%
}%
\fi

\usepackage[frozencache,cachedir=minted-cache]{minted}
\usepackage{etoolbox}

\makeatletter
\ifcsdef{minted@optlistcl@quote}
{
\ifwindows
  \renewcommand{\minted@optlistcl@quote}[2]{%
    \ifstrempty{#2}{\detokenize{#1}}{\detokenize{#1="#2"}}}
\else
  \renewcommand{\minted@optlistcl@quote}[2]{%
    \ifstrempty{#2}{\detokenize{#1}}{\detokenize{#1='#2'}}}
\fi

\newcommand{\minted@def@optcl@novalue}[2]{%
  \define@booleankey{minted@opt@g}{#1}%
    {\minted@addto@optlistcl{\minted@optlistcl@g}{#2}{}%
     \@namedef{minted@opt@g:#1}{true}}
    {\@namedef{minted@opt@g:#1}{false}}
  \define@booleankey{minted@opt@g@i}{#1}%
    {\minted@addto@optlistcl{\minted@optlistcl@g@i}{#2}{}%
     \@namedef{minted@opt@g@i:#1}{true}}
    {\@namedef{minted@opt@g@i:#1}{false}}
  \define@booleankey{minted@opt@lang}{#1}%
    {\minted@addto@optlistcl@lang{minted@optlistcl@lang\minted@lang}{#2}{}%
     \@namedef{minted@opt@lang\minted@lang:#1}{true}}
    {\@namedef{minted@opt@lang\minted@lang:#1}{false}}
  \define@booleankey{minted@opt@lang@i}{#1}%
    {\minted@addto@optlistcl@lang{minted@optlistcl@lang\minted@lang @i}{#2}{}%
     \@namedef{minted@opt@lang\minted@lang @i:#1}{true}}
    {\@namedef{minted@opt@lang\minted@lang @i:#1}{false}}
  \define@booleankey{minted@opt@cmd}{#1}%
      {\minted@addto@optlistcl{\minted@optlistcl@cmd}{#2}{}%
        \@namedef{minted@opt@cmd:#1}{true}}
      {\@namedef{minted@opt@cmd:#1}{false}}
}
\minted@def@optcl@novalue{customlexer}{-x}
}
{
}
\makeatother

\usepackage{tikz}
\usetikzlibrary{arrows}
\usetikzlibrary{shapes}

\tikzset{
  circledstyle/.style={
    shape=circle,
    #1,
    font=\tt\small,
    inner sep=0pt,
    outer sep=0pt,
    minimum size=1.2em,
    text=black
  }
}

\DeclareRobustCommand{\circledbase}[3][]{%
    \tikz[baseline=(char.base)]{\node[circledstyle, fill=#2] (char) {#3\strut};}%
}

\usepackage[norefs,nocites]{refcheck}

\makeatletter
\newcommand{\refcheckize}[1]{%
  \expandafter\let\csname @@\string#1\endcsname#1%
  \expandafter\DeclareRobustCommand\csname relax\string#1\endcsname[1]{%
    \csname @@\string#1\endcsname{##1}\wrtusdrf{##1}}%
  \expandafter\let\expandafter#1\csname relax\string#1\endcsname
}
\makeatother

\refcheckize{\autoref}

%% file: tex/acm.tex
\ifx\acmversion\acmversionjournal
  \acmJournal{PACMPL}
  \acmVolume{1}
  \acmNumber{1}
  \acmArticle{1}
  \acmYear{2017}
  \acmMonth{1}
  \acmDOI{10.1145/nnnnnnn.nnnnnnn}
  \startPage{1}
\else
  \acmConference[ASPLOS'26]{International Conference on Architectural Support for Programming Languages and Operating Systems}{March 2026}{Pittsburgh, USA}
  \acmYear{2026}
  \acmISBN{978-x-xxxx-xxxx-x/YY/MM}
  \acmDOI{10.1145/nnnnnnn.nnnnnnn}
  \startPage{1}
\fi